\documentclass{article}
\usepackage{graphicx}
\usepackage{spconf,amsmath,bm,cite,graphicx} %amssymb
\usepackage{makecell,booktabs,arydshln}

\usepackage{color}
\usepackage{multirow}
\usepackage{etoolbox,siunitx}
\robustify\bfseries
\usepackage{balance}

\usepackage{algorithm}
\usepackage{algpseudocode}
\algrenewcommand\algorithmicindent{0.9em}%
\makeatletter
\def\adl@drawiv#1#2#3{%
        \hskip.5\tabcolsep
        \xleaders#3{#2.5\@tempdimb #1{1}#2.5\@tempdimb}%
                #2\z@ plus1fil minus1fil\relax
        \hskip.5\tabcolsep}
        
\newcommand{\cdashlinelr}[1]{%
  \noalign{\vskip\aboverulesep
           \global\let\@dashdrawstore\adl@draw
           \global\let\adl@draw\adl@drawiv}
  \cdashline{#1}
  \noalign{\global\let\adl@draw\@dashdrawstore
           \vskip\belowrulesep}}
           
\newcommand{\myhdashline}{%
  \noalign{\vskip\aboverulesep
           \global\let\@dashdrawstore\adl@draw
           \global\let\adl@draw\adl@drawiv}
  \hdashline
  \noalign{\global\let\adl@draw\@dashdrawstore
           \vskip\belowrulesep}}
\makeatother

\title{Extended Graph Temporal Classification\\for Multi-Speaker End-to-End ASR}
\name{Xuankai Chang$^{1,2}$, Niko Moritz$^1$, Takaaki Hori$^1$, Shinji Watanabe$^2$, Jonathan Le Roux$^1$ \thanks{Work performed while X. Chang was an intern at MERL.}}
\address{
    $^1$Mitsubishi Electric Research Laboratories (MERL), Cambridge, MA, USA\\
    $^2$Language Technologies Institute, Carnegie Mellon University, Pittsburgh, PA, USA
}

\begin{document}
\ninept
\setlength{\abovedisplayskip}{4.5pt}
\setlength{\belowdisplayskip}{4.5pt}
\allowdisplaybreaks

\maketitle

\begin{abstract}

Graph-based temporal classification (GTC), a generalized form of the connectionist temporal classification loss, was recently proposed to improve automatic speech recognition (ASR) systems using graph-based supervision.
For example, GTC was first used to encode an N-best list of pseudo-label sequences into a graph for semi-supervised learning.
In this paper, we propose an extension of GTC to model the posteriors of both labels and label transitions
by a neural network, which can be applied to a wider range of tasks.
As an example application, we use the extended GTC (GTC-e) for the multi-speaker speech recognition task. The transcriptions and speaker information of multi-speaker speech are represented by a graph
, where the speaker information is associated with the transitions and ASR outputs with the nodes. 
Using GTC-e, multi-speaker ASR modelling becomes very similar to single-speaker ASR modeling, in that tokens by multiple speakers are recognized as a single merged sequence in chronological order.
For evaluation, we perform experiments on a simulated multi-speaker speech dataset derived from LibriSpeech, obtaining promising results with performance close to classical benchmarks for the task.
\end{abstract}
\section{Introduction}
\label{sec:intro}

In recent years, dramatic progress has been achieved in automatic speech recognition (ASR), in particular thanks to the exploration of neural network architectures that improve the robustness and generalization ability of ASR models \cite{qian2016very, graves2013speech, VaswaniSPUJGKP17, gulati2020conformer, guo2021recent}. The rise of end-to-end ASR models has simplified ASR architecture with a single neural network, with frameworks such as the connectionist temporal classification (CTC) \cite{GravesFGS06}, attention-based encoder-decoder model \cite{chan2015listen,kim2017joint,watanabe2017hybrid}, and the RNN-Transducer model \cite{graves2012sequence}.

\begin{figure}
    \vspace{-12pt}
    \centering
    \includegraphics[width=\linewidth]{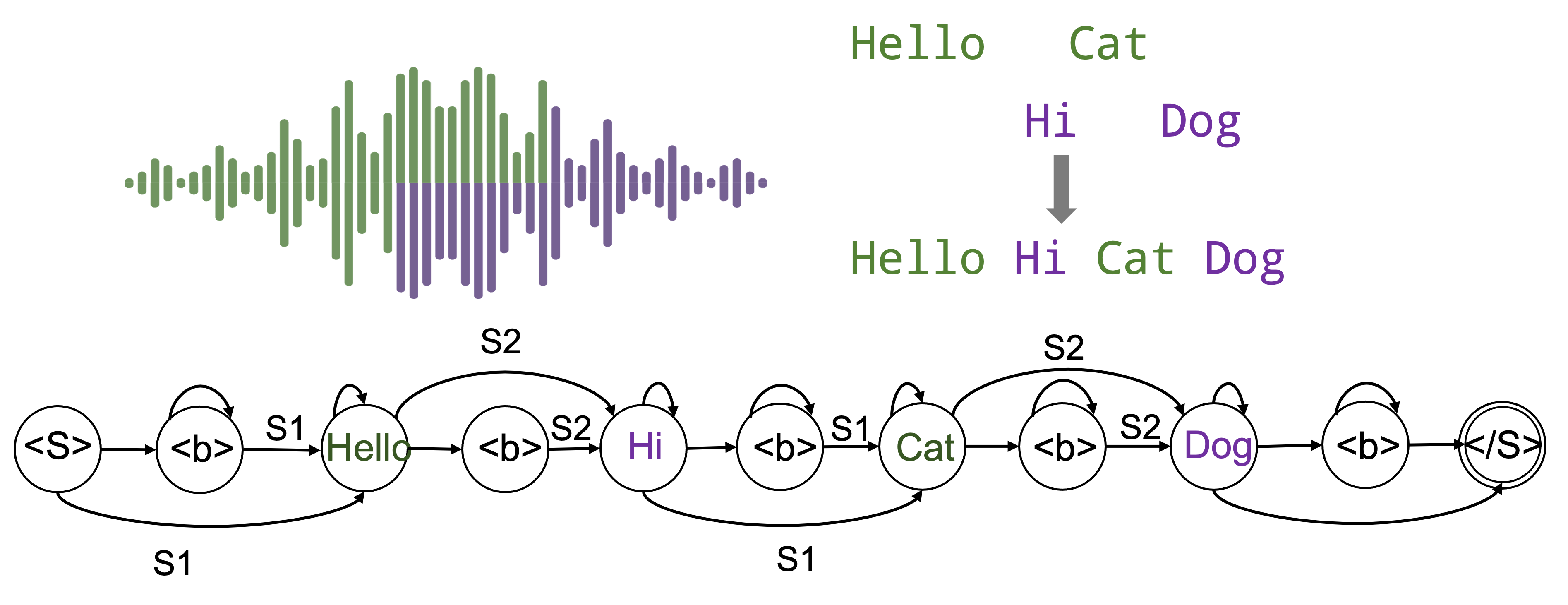}
    \caption{Illustration of a GTC-e graph for multi-speaker ASR. In the graph, the nodes represent the tokens (words) from the transcriptions. The edges indicate the speaker transitions.}
    \label{fig:wfst}
    \vspace{-12pt}
\end{figure}

Graph modeling has traditionally been used in ASR for decades.
For example, in 
hidden Markov model (HMM) based systems, a weighted finite-state transducer (WFST) is used to combine several modules together including a pronunciation lexicon, context-dependencies, and a language model (LM)~\cite{mohri2002weighted,hori2007efficient}. Recently, researchers proposed to use graph representations in the loss function for training deep neural networks \cite{hannun2020differentiable}. In \cite{moritz2021semi}, a new loss function, called graph-based temporal classification (GTC), was proposed as a generalization of CTC to handle sequence-to-sequence problems. GTC can take 
graph-based supervisory information 
as an input 
to describe all possible alignments between an input sequence and an output sequence, for learning the best possible alignment from the training data.
As an example of application, GTC was used to boost ASR performance via semi-supervised training \cite{Lamel2002lightlySupASR,Huang2013SemisupervisedGA} by using an N-best list of ASR hypotheses that is converted into a graph representation to train an ASR model using unlabeled data. 
However, in the original GTC, only posterior probabilities of the ASR labels are trained, and trainable label transitions are not considered. Extending GTC to handle label transitions would allow us to model further information regarding the labels.  For example, in a multi-speaker speech recognition scenario, where some overlap between the speech signals of multiple speakers is considered, we could use the transition weights to model speaker predictions that are aligned with the ASR label predictions at frame level, such that when an ASR label is predicted we can also detect if it belongs to a specific speaker.
Such a graph is illustrated in Fig.~\ref{fig:wfst}.

In the last few years, several multi-speaker end-to-end ASR models have been proposed. In \cite{Seki2018ACL07,chang2019end}, permutation invariant training (PIT) \cite{Hershey2016ICASSP03,Isik2016Interspeech09,yu2017permutation} was used to compute the loss by choosing the hypothesis-reference assignment with minimum loss. In \cite{kanda2020serialized}, an attention-based encoder-decoder is trained to generate the hypothesis sequences of different speakers in a predefined order based on heuristic information, a technique called serialized output training (SOT). In \cite{shi2020sequence,guo2021multi}, the model is trained to predict the hypothesis sequence of one speaker in each iteration while utilizing information about the previous speakers' hypotheses as additional input.
These existing multi-speaker end-to-end ASR models, which have showed promising results, all share a common characteristic in the way that the predictions can be divided at the level of a whole utterance for each speaker. For example, in the PIT-based methods, label sequences for different speakers are supposed to be output at different output heads, while in the SOT-/conditional-based models, the prediction of the sequence for a speaker can only start when the sequence of the previous speaker completes.

In contrast to previous works, in this paper, the multi-speaker ASR problem is not implicitly regarded as a source separation problem using separate output layers for each speaker or cascaded processes to recognize each speaker one after another. Instead, the prediction of ASR labels of multiple speakers is regarded as a sequence of acoustic events irrespective of the source shown as in Fig.~\ref{fig:wfst}, and the belonging to a source is predicted separately to distinguish if an ASR label was uttered by a given speaker.
We propose to use an extended GTC (GTC-e) loss to accomplish this, which allows us to train two separate predictions, one for the speakers and one for the ASR outputs, that are aligned at the frame level.
In order to exploit the speaker predictions efficiently during decoding, we also modify an existing frame-synchronous beam search algorithm to adapt it to GTC-e.
The proposed model is evaluated on a multi-speaker end-to-end ASR task based on the LibriMix data, including various degrees of overlap between speakers.
To the best of our knowledge, this is the first work to address multi-speaker ASR 
by considering the ASR outputs of multiple speakers as a sequence of intermingled events with a chronologically meaningful ordering.

\section{Extended GTC (GTC-e)}
\label{sec:gtc}

In this section, we describe the extended GTC loss function. For the convenience of understanding, we mostly follow the notations in the previous GTC study \cite{moritz2021semi}.

GTC was proposed as a loss function to address sequence-to-sequence problems. We assume the input of the neural network is a sequence denoted as $X=(x_1, \dots, x_{L})$, where $L$ stands for the length. The output is a sequence of length $T$, $Y=(\mathbf{y}^1, \dots, \mathbf{y}^T)$, where $\mathbf{y}^t$ denotes the posterior probability distribution over an alphabet $\mathcal{U}$, and the $k$-th class's probability is denoted by $y_k^t$. We use $\mathcal{G}$ to refer to a graph constructed from references. Then the GTC function computes the posterior probability for graph $\mathcal{G}$ by summing over all alignment sequences in $\mathcal{G}$:
\begin{align}
    p(\mathcal{G}|X) = \sum_{\pi \in \mathcal{S}(\mathcal{G},T)} p(\pi | X), \label{eq:gtc}
\end{align}
where $\mathcal{S}$ represents a search function that unfolds $\mathcal{G}$ to all possible node sequences of length $T$ (not counting non-emitting start and end nodes), $\pi$ denotes a single sequence of nodes, and $p(\pi|X)$ is the posterior probability for $\pi$ given feature sequence $X$. 
The loss function is defined as the following negative log likelihood:
\begin{align}
    \mathcal{L} = -\ln{p(\mathcal{G}|X)}. 
\label{eq:loss}
\end{align}

Following \cite{moritz2021semi}, we index the nodes of graph $\mathcal{G}$ using $g=0,\dots,G+1$, sorting them in a breadth-first search manner from $0$ (non-emitting start node) to $G+1$ (non-emitting end node). We denote by $l(g) \in \mathcal{U}$ the output symbol observed at node $g$, and by $W_{(g,g')}$ a deterministic transition weight on edge ($g$, $g'$). In addition, we denote by $\pi_{t:t'}=(\pi_t,\dots,\pi_{t'})$ the node sub-sequence of $\pi$ from time index $t$ to $t'$. Note that $\pi_0$ and $\pi_{T+1}$ correspond to the non-emitting start and end nodes $0$ and $G+1$, respectively.

We modify GTC such that the neural network can generate an additional posterior probability distribution, $\mathbf{\omega}^t_{I(g, g')}$, representing a transition weight on edge $(g, g')$ at time $t$, where $I(g, g')\in\mathcal{I}$ and $\mathcal{I}$ is the index set of all possible transitions. The posterior probabilities are obtained as the output of a softmax.
The forward probability, $\alpha_t(g)$, represents the total probability at time $t$ of the sub-graph $\mathcal{G}_{0:g}$  of $\mathcal{G}$ containing all paths from node $0$ to node $g$. It can be computed for $g=1,\dots,G$ using
\begin{align}
    \alpha_t (g) = \sum_{\substack{\pi \in \mathcal{S}(\mathcal{G},T):\\ \pi_{0:t} \in \mathcal{S}(\mathcal{G}_{0:g},t)}} \prod_{\tau=1}^t W_{\pi_{\tau-1},\pi_\tau}  \omega^{\tau}_{I(\pi_{\tau-1},\pi_\tau)}  y_{l(\pi_{\tau})}^{\tau}. \label{eq:grapctc_fw_probs_ext}
\end{align}
Note that $\alpha_0(g)$ equals $1$ if $g$ corresponds to the start node and it equals $0$ otherwise.
The backward probability $\beta_t (g)$ is computed similarly, using
\begin{align}
    \label{eq:grapctc_bw_probs_ext}
    \resizebox{.91\hsize}{!}{
        $
        \displaystyle \beta_t (g) =\!\!\!\!\!\!\!\!\!\!\!\!\!\!\! \sum_{
            \substack{\pi \in \mathcal{S}(\mathcal{G},T):\\ \pi_{t:T+1} \in \mathcal{S}(\mathcal{G}_{g:G+1},T-t+1)}
        }\!\!\!\!\!\!\!\!\!\!\!\!\!\!\! \left[ 
            y_{l(\pi_{T})}^{T} \prod_{\tau=t}^{T-1} W_{\pi_{\tau},\pi_{\tau+1}} \omega^{\tau+1}_{I(\pi_{\tau},\pi_{\tau+1})} y_{l(\pi_{\tau})}^{\tau}
        \right],
        $
    }
\end{align}
where $\mathcal{G}_{g:G+1}$ denotes the sub-graph of $\mathcal{G}$ containing all paths from node $g$ to node $G+1$. Similar to GTC or CTC, the computation of $\alpha$ and $\beta$ can be efficiently performed using the forward-backward algorithm.

The network is optimized by gradient descent. The gradients of the loss with respect to the label posteriors $y_{k}^{t}$ and to the corresponding unnormalized network outputs $u_k^t$ before the softmax is applied, for any symbol $k \in \mathcal{U}$, can be obtained in the same way as in CTC and GTC, where the key idea is to express the probability function $p(\mathcal{G}|X)$ at $t$ using the forward and backward variables:
\begin{align}
 p(\mathcal{G}|X) = \sum_{g \in \mathcal{G}} \frac{\alpha_t(g) \beta_t(g)}{y^{t}_{l(g)}}.
\label{eq:fw_bw}
\end{align}

The derivation of the gradient of the loss with respect to the network outputs for the transition probabilities $w_i^t$, for a transition $i \in \mathcal{I}$, is similar but with some important differences. Here, the key is to express $p(\mathcal{G}|X)$ at $t$ as
\begin{equation}
\label{eq:fw_bw_ext_weights}
    p(\mathcal{G}|X) = \sum_{(g,g') \in \mathcal{G}} \alpha_{t-1}(g) W_{g,g'} \omega^{t}_{I(g,g')}  \beta_t(g').
\end{equation}
The derivative of $p(\mathcal{G}|X)$ with respect to the transition probabilities $\omega_i^t$ can then be written as
\begin{equation}
\label{eq:deriv_pGX_y_ext}
    \frac{\partial  p(\mathcal{G}|X)}{\partial \omega^{t}_{i}}  = \sum_{(g,g') \in \operatorname{\Phi}(\mathcal{G},i)}
    \alpha_{t-1}(g) W_{g,g'} \beta_t(g')  ,
\end{equation}
where $\operatorname{\Phi}(\mathcal{G},i) = \{(g,g') \in \mathcal{G} : I(g,g') = i \}$ denotes the set of edges in $\mathcal{G}$ that correspond to transition $i$. 
To backpropagate the gradients through the softmax function of $w_i^t$, we need the derivative with respect to the unnormalized network outputs $h_i^t$ before the softmax is applied, which is 
\begin{equation}
\label{eq:deriv_pGX_u_ext}
    -\frac{\partial \ln{{p}(\mathcal{G}|X)}}{\partial h_i^t} = - \sum_{i' \in \mathcal{I}} \frac{\partial \ln{{p}(\mathcal{G}|X)}}{\partial \omega_{i'}^{t}} \frac{\partial \omega_{i'}^t}{\partial h_i^t}.
\end{equation}
The gradients for the transition weights are derived by substituting (\ref{eq:deriv_pGX_y_ext}) and the derivative of the softmax function $ \partial \omega_{i'}^t / \partial h_i^t = \omega_{i'}^t \delta_{i i'}  -  \omega_{i'}^t \omega_k^t $ into (\ref{eq:deriv_pGX_u_ext}):
\begin{align}
   \hspace{-.2cm} -\frac{\partial \ln{ p(\mathcal{G}|X)} }{\partial h_i^t}\! =\! \omega_i^t \!-\!  \frac{\omega_i^t}{{p}(\mathcal{G}|X)} \sum_{(g,g') \in \operatorname{\Phi}(\mathcal{G},i)}\!\!\!\!\!\!\!\! \alpha_{t-1}(g) W_{g,g'} \beta_t(g').
\end{align}
We used the fact that
\begin{align}
    -\sum_{i' \in \mathcal{I}} \frac{\partial \ln{{p}(\mathcal{G}|X)}}{\partial \omega_{i'}^t} \omega_{i'}^t \delta_{i i'} &= -\frac{\partial \ln{{p}(\mathcal{G}|X)}}{\partial \omega_{i}^t} \omega_{i}^t ,  \nonumber \\
    &  \hspace{-2cm}= -\frac{\omega_i^t}{{p}(\mathcal{G}|X)} \sum_{(g,g') \in \operatorname{\Phi}(\mathcal{G},i)} \alpha_{t-1}(g) W_{g,g'} \beta_t(g') ,
\end{align}
and that
\begin{align}
    &  \hspace{-1cm}\sum_{i' \in \mathcal{I}} \frac{\partial \ln{{p}(\mathcal{G}|X)}}{\partial \omega_{i'}^t} \omega_{i'}^t \omega_{i}^t \nonumber  \nonumber \\
    =& \sum_{i' \in \mathcal{I}} \frac{\omega_{i'}^t \omega_{i}^t}{{p}(\mathcal{G}|X)} \sum_{(g,g') \in \operatorname{\Phi}(\mathcal{G},i')} \alpha_{t-1}(g) W_{g,g'} \beta_t(g') ,  \nonumber \\
    =&  \frac{ \omega_{i}^t}{{p}(\mathcal{G}|X)} \sum_{i' \in \mathcal{I}} \sum_{(g,g') \in \operatorname{\Phi}(\mathcal{G},i')} \alpha_{t-1}(g) W_{g,g'} \omega_{i'}^t \beta_t(g') ,  \nonumber \\
    =&  \frac{ \omega_{i}^t}{{p}(\mathcal{G}|X)} \sum_{(g,g') \in \mathcal{G}} \alpha_{t-1}(g) W_{g,g'} \omega_{I(g,g')}^t \beta_t(g') ,  \nonumber \\
    =& \frac{\omega_{i}^t}{{p}(\mathcal{G}|X)} {p}(\mathcal{G}|X) = \omega_i^t.
\end{align}

For efficiency reason, we implemented the GTC objective in CUDA as an extension for PyTorch.

\section{Multi-speaker ASR and Beam Search}
\label{sec:beamsearch}

We apply the extended GTC approach to multi-speaker ASR, which is considered as a challenging task in the field of speech processing. One of the main difficulties of multi-speaker ASR stems from the necessity
to find a way to train a network that will be able to reliably group tokens from the same speaker together. Most existing approaches attempt to handle
this problem either by splitting the speakers across multiple outputs \cite{yu2017permutation,chang2019end} or by making predictions sequentially speaker by speaker \cite{kanda2020serialized,shi2020sequence,guo2021multi}. The ambiguity in how to assign a given output to a given reference at training time is typically broken either by using permutation invariant training or by using an arbitrary criterion such as assigning an output to the speaker with highest energy or with the earliest onset. We here take a completely different approach, motivated by our noticing that a graph can be a good representation for overlapped speech, since it can represent the tokens at each node while the speaker identity can also be labeled at each edge. More specifically, given the transcriptions of all the speakers in an overlapped speech, we can convert them to a sequence of
chronologically ordered linguistic tokens where each token has a speaker identity. The temporal alignment of tokens can be acquired by performing CTC alignment on each isolated clean speech, which is like a sequence of sparse spikes, as shown in Fig.~\ref{fig:speaker_probability}, and merging them based on their time occurrence. Note here that this assumes that the activation period of linguistic tokens from different speakers are not completely the same. In practice, this condition is often satisfied, although overlaps do occur in some small percentage of frames. Based on this, we can construct a graph for multi-speaker ASR for each overlapped speech mixture. We show a simple example graph in Fig.~\ref{fig:wfst}.
In this setup, the alphabet $\mathcal{U}$ for the node labels consists of all the ASR tokens, and the set of transitions $\mathcal{I}$ consists of the speaker indices up to the maximum number of speakers.

As in GTC\cite{moritz2021semi}, we can apply a beam search algorithm during decoding. Since the output of GTC-e contains tokens from multiple speakers, we need to make modifications to the existing time-synchronous prefix beam search algorithm \cite{hannun2014first,moritz2019streaming}. The modified beam search is shown in Algorithm~\ref{alg:beamsearch}. The main modifications are three fold. First, we apply the speaker transition probability in the score computation. Second, when expanding the prefixes, we need to consider all possible speakers. Third, when computing the LM scores of a prefix, we need to consider the sub-sequences of different speakers separately. 
\begin{algorithm}[t]
    \caption{The modified time-synchronous prefix beam search for extended GTC. We use $A_{\text{prev}}$ to store every prefix $l$ at every time step. We denote the alphabet by $\mathcal{U}$ and number of speakers by ${S}$. We denote the symbol posterior by $p(\cdot)$ and the speaker transition posterior by $p^\omega(\cdot)$.}
    \label{alg:beamsearch}
    \begin{algorithmic}[1]
        \State $\ell \leftarrow \left(\left(\left\langle sos \right\rangle,0 \right),\right)$
        \State $p_b(\ell) \leftarrow 1$, $p_{nb}(\ell) \leftarrow 0$
        \State $A_{\text{prev}} \leftarrow \{\ell\}$
        \For{t=1,\dots,T}
            \State $A_{\text{next}} \leftarrow \{\}$
            \For{$\ell$ \textbf{in} $A_\text{prev}$}
              \For{$c$ \textbf{in} $\mathcal{U}$}
                  \If{$c = \text{blank}$}
                      \State{
                          $\begin{aligned}
                          p_b(\ell) &\leftarrow p(\text{blank}; x_t) p^\omega(\text{blank}; x_t) ( p_b(\ell; x_{1:t-1}) +  \\ 
                          &\hspace{3.95cm}p_{nb}(\ell; x_{1:t-1}) )
                          \end{aligned}$
                      }
                      \State add $\ell$ to $A_\text{next}$
                  \Else
                      \For{$s = 1, \dots, S$} \Comment{Loop over speaker index}
                          \State $\ell^{+} \leftarrow \text{ append } \left(c, s\right) \text{ to } \ell $
                          \If{$\left(c,s\right) = \ell_\text{end}$}
                              \State $p_{nb}(\ell^{+}; x_{1:t}) \leftarrow p(c; x_t) p_b(\ell;x_{1:t-1}) p^{\omega}(s; x_t)$
                              \State $p_{nb}(\ell; x_{1:t}) \leftarrow p(c; x_t) p_{nb}(\ell;x_{1:t-1}) p^{\omega}(s; x_t)$
                          \Else
                              \State{
                              $\begin{aligned}
                              p_{nb}(\ell^+; x_{1:t}) &\leftarrow  p(c; x_t)  (p_b(\ell; x_{1:t-1}) +  \\ 
                              &\hspace{.8cm}  p_{nb}(\ell; x_{1:t-1})) p^{\omega}(s; x_t) 
                              \end{aligned}$
                              }
                          \EndIf
                          \If{$\ell^+ \text{\bf{not in}} A_{\text{prev}}$}
                              \State{
                              $\begin{aligned}
                                  p_{b}(\ell^+; x_{1:t}) &\leftarrow  p(\text{blank}; x_t)  (p_b(\ell^+; x_{1:t-1}) + \\ 
                              &\hspace{.5cm}p_{nb}(\ell^+; x_{1:t-1})) p^{\omega}(\text{blank}; x_t)
                              \end{aligned}$
                              }
                              \State{
                              $\begin{aligned}
                              p_{nb}(\ell^+; x_{1:t}) \leftarrow p(c; x_t) & p_{nb}(\ell^+; x_{1:t-1}) \cdot \\
                              &\hspace{1.5cm} p^{\omega}(s; x_t)
                              \end{aligned}$
                              }

                          \EndIf
                          \State add $\ell^+$ to $A_{\text{next}}$
                      \EndFor
                  \EndIf
              \EndFor
            \EndFor
            \State $A_{\text{prev}} \leftarrow k \text{ most probable prefixes in } A_{\text{next}}$ \Comment{Track the LM scores of different speakers separately.}
        \EndFor
    \end{algorithmic}
    \vspace{-1pt}
\end{algorithm}

\section{Experiments}
\label{sec:experiments}

\subsection{Setup}
\label{ssec:setup}

We carried out multi-speaker end-to-end speech recognition experiments using the LibriMix~\cite{cosentino2020librimix} dataset as well as data derived from it. LibriMix contains multi-speaker overlapped speech simulated by mixing utterances randomly chosen from different speakers in the LibriSpeech corpus~\cite{librispeech}. For fast adaption, we use the 2-speaker train\_clean\_100 subset of LibriMix. The original LibriMix dataset generates fully overlapped speech by default, which means that one utterance is 100\% interfered by the other (assuming they have the same length). However, in realistic conditions, the overlap ratio is usually small \cite{cetin1analysis,chen2020continuous}. To simulate such conditions, we use the same utterance selections and signal to noise ratio (SNR) as in LibriMix with smaller overlapping ratios of $0\%$ and $40\%$ to generate additional training data subsets.

\begin{figure*}[t]
    \centering
    \includegraphics[width=\linewidth]{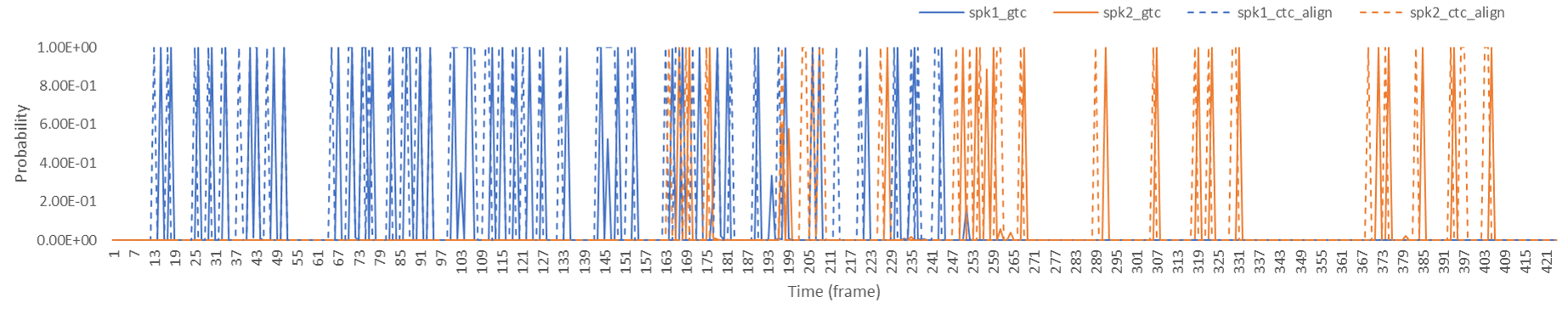}
    \vspace{-20 pt}
    \caption{An example of speaker transition posterior predicted by GTC. The input 2-speaker utterance's overlap ratio is about $40\%$. The figure shows the predicted (solid line) and ground truth (dashed line) activations.}
    \label{fig:speaker_probability}
    \vspace{-20 pt}
\end{figure*}

For labels, we use the linguistic token sequence of all the speakers in the mixture. First, we generate the token alignments given each source utterance based on the Viterbi alignment of CTC, which indicates the rough activation time of every token. Then we combine the alignments of two speakers by ordering the tokens monotonically along the time axis. In order to reduce the concurrent activations of tokens from different speakers, we make use of byte pair encodings (BPE) as our token units. In our experiments, we use the BPE model with 5000 tokens trained on LibriSpeech data. The concurrent activations of tokens for two speakers are relatively rare, at the rate of $6\%$ and $2\%$ on fully and $40\%$ overlapping ratio training subsets respectively. 
When these concurrent activations occur, we 
use a predefined order which makes the label from the speaker with highest energy over the whole utterance come first (allowing multiple permutations in the label graph will be considered in future work).

For ASR models, we simply reused the encoder architecture in PIT-based multi-speaker end-to-end speech recognition models, for the details of which we shall refer the reader to \cite{chang2019end}.
In the model, there are 2 CNN blocks to encode the input acoustic feature, followed by 2 sub-networks each of which has 4 Transformer layers to extract the token and speaker information, respectively. Then 8 shared Transformer layers are used to convert each of the two sequences to some representation. For the two output sequences, one is regarded as token hidden representation and the other one is regarded as speaker prediction.
We use a normal single-speaker ASR model trained with CTC (single-speaker CTC)
and the original end-to-end PIT-ASR model \cite{chang2019end} trained with CTC loss only (PIT-CTC) as our baselines.

\subsection{Greedy search results}
\label{ssec:exp1}

\begin{table}[t]
    \centering
    \sisetup{table-format=2.1,round-mode=places,round-precision=1,table-number-alignment = center,detect-weight=true,detect-inline-weight=math}
    \caption{WER(\%) comparison between baselines and the GTC-e model using greedy search decoding.}
    \resizebox{\linewidth}{!}{\begin{tabular}{lSSSSSSSS}
        \toprule
        & \multicolumn{2}{c}{0\% overlap} & \multicolumn{2}{c}{20\% overlap} & \multicolumn{2}{c}{40\% overlap} & \multicolumn{2}{c}{Full overlap} \\
        \cmidrule(lr){2-3} \cmidrule(lr){4-5} \cmidrule(lr){6-7} \cmidrule(lr){8-9} 
         Model & {dev} & {test} & {dev} & {test}& {dev} & {test}& {dev} & {test} \\
        \midrule
        single-speaker CTC & 34.6 & 34.1 & 37.4 & 37.0 & 45.9 & 45.3 & 76.3 & 75.9 \\
        PIT-CTC & 18.8 & 19.2 & 19.9 & 22.3 & 22.9 & 23.5 & 32.9 & 33.8 \\
        GTC-e & 20.5 & 21.1 & 22.6 & 23.3 & 26.3 & 27.3 & 44.6 & 45.8 \\
        \bottomrule
    \end{tabular}}
    \label{tab:greedy}
    \vspace{-15pt}
\end{table}

\begin{table}[t]
    \centering
    \sisetup{table-format=2.1,round-mode=places,round-precision=1,table-number-alignment = center,detect-weight=true,detect-inline-weight=math}
    \caption{Oracle TER(\%) comparison between PIT-CTC and GTC-e.}
    \resizebox{\linewidth}{!}{\begin{tabular}{lSSSSSSSSSS}
        \toprule
         & \multicolumn{2}{c}{0\% overlap} & \multicolumn{2}{c}{20\% overlap} & \multicolumn{2}{c}{40\% overlap} & \multicolumn{2}{c}{Full overlap} & \multicolumn{2}{c}{Average}  \\
        \cmidrule(lr){2-3} \cmidrule(lr){4-5} \cmidrule(lr){6-7} \cmidrule(lr){8-9} \cmidrule(lr){10-11} 
        Model & {dev} & {test} & {dev} & {test}& {dev} & {test}& {dev} & {test} & {dev} & {test} \\ 
        \midrule
        PIT-CTC & 18.5 & 18.4 & 19.4 & 19.5 & 22.0 & 22.4 & 30.1 & 30.9 & 22.5 & 22.8  \\
        GTC-e & 19.8 & 20.1 & 21.1 & 21.4 & 24.1 & 24.6 & 33.4 & 33.9 & 24.6 & 25.0 \\
        \bottomrule
    \end{tabular}}
    \label{tab:oracle}
    \vspace{-12pt}
\end{table}

In this section, we describe the ASR performance of the baselines and the proposed GTC-e model using greedy search decoding. The word error rates (WERs) are shown in Table~\ref{tab:greedy}. From the table, we can see that the proposed model is better than the normal ASR model. Our proposed model also achieves a performance close to the PIT-CTC model, especially in the low-overlap ratio cases (0\%, 20\%, 40\%). Note that although our model predicts the speaker indices, there exists speaker prediction errors. We further check the oracle token error rates (TER) of PIT-CTC and GTC-e, by only comparing the tokens from all output sequences against all reference sequences, regardless of speaker assignment. As shown in Table~\ref{tab:oracle}, we obtain averaged test TERs for PIT-CTC and GTC-e of $22.8\%$ and $25.0\%$ respectively, from which we can tell that the token recognition performance is comparable. It indicates that we should consider how to improve the speaker prediction in the next step.

We also show an example of CTC ground truth token alignment together with the speaker transition posterior predictions by our model in Fig.~\ref{fig:speaker_probability}. From the figure, we can see that our GTC-e model can accurately predict the activations of most tokens.

\subsection{Beam Search Results}
\label{ssec:exp2}
We here present the ASR performance of beam search decoding, shown in Table~\ref{tab:beamsearch}. For the language model, we use a 16-layer Transformer-based LM trained on full LibriSpeech data with external text. The beam size of GTC-e is set to 40, while that of PIT-CTC is cut to half to keep the average beam size of every speaker the same. With the beam search, the word error rates are greatly improved. Our approach obtains promising results which are close to the PIT-CTC baseline, albeit with a slightly worse WER. In addition to the average WERs, the WERs for each speaker are also shown in Table~\ref{tab:beamsearch-ind}, confirming that the model is not biased towards a particular speaker output.

\begin{table}[t]
    \centering
    \sisetup{table-format=2.1,round-mode=places,round-precision=1,table-number-alignment = center,detect-weight=true,detect-inline-weight=math}
    \caption{WER(\%) comparison between PIT-CTC and GTC-e using beam search decoding.}
    \resizebox{\linewidth}{!}{\begin{tabular}{lSSSSSSSS}
        \toprule
         & \multicolumn{2}{c}{0\% overlap} & \multicolumn{2}{c}{20\% overlap} & \multicolumn{2}{c}{40\% overlap} & \multicolumn{2}{c}{Full overlap} \\
        \cmidrule(lr){2-3} \cmidrule(lr){4-5} \cmidrule(lr){6-7} \cmidrule(lr){8-9} 
        Model & {dev} & {test} & {dev} & {test}& {dev} & {test}& {dev} & {test} \\ 
        \midrule
        PIT-CTC & 11.7 & 12.4 & 12.6 & 13.4 & 16.3 & 18.1 & 24.0 & 26.3 \\
        GTC-e & 14.8 & 15.5 & 16.5 & 17.2 & 19.5 & 20.4 & 32.7 & 33.7 \\
        \bottomrule
    \end{tabular}}
    \label{tab:beamsearch}
    \vspace{-15pt}
\end{table}

\begin{table}[t]
    \centering
    \sisetup{table-format=2.1,round-mode=places,round-precision=1,table-number-alignment = center,detect-weight=true,detect-inline-weight=math}
    \caption{WER(\%) for each speaker with GTC-e using beam search decoding.}
    \resizebox{\linewidth}{!}{\begin{tabular}{lSSSSSSSS}
        \toprule
         & \multicolumn{2}{c}{0\% overlap} & \multicolumn{2}{c}{20\% overlap} & \multicolumn{2}{c}{40\% overlap} & \multicolumn{2}{c}{Full overlap} \\
        \cmidrule(lr){2-3} \cmidrule(lr){4-5} \cmidrule(lr){6-7} \cmidrule(lr){8-9} 
        Speaker & {dev} & {test} & {dev} & {test}& {dev} & {test}& {dev} & {test} \\ 
        \midrule
        spk1 & {15.0} & {15.1} & {17.0} & {17.3} & {20.6} & {21.1} & {33.0} & {33.7} \\
        spk2 & {14.7} & {15.7} &  {15.9} &  {17.1} &  {18.4} &  {19.7} & {32.3} &{33.7} \\
        \bottomrule
    \end{tabular}}
    \label{tab:beamsearch-ind}
    \vspace{-15pt}
\end{table}

\section{Conclusion}
\label{sec:conclusion}

In this paper, we proposed GTC-e, an extension of the graph-based temporal classification method using neural networks to predict posterior probabilities for both labels and label transitions. This extended GTC framework opens the way to a wider range of applications. 
As an example application, we explored the use of GTC-e for multi-speaker end-to-end ASR, 
a notably challenging task, leading to a multi-speaker ASR system that transcribes speech in a very similar way to single-speaker ASR. 
We have performed preliminary experiments on the LibriMix 2-speaker dataset, showing promising results demonstrating the feasibility of the approach. 
In future work, we will explore other applications of GTC-e and investigate ways to improve the performance of extended GTC on multi-speaker ASR by using new model architectures 
and by exploring label and speaker permutations in the graph to allow for more flexible alignments.

\balance
\bibliographystyle{IEEEtran}
\bibliography{refs}

\end{document}